# Human Communication Systems Evolve by Cultural Selection

NICOLAS FAY, University of Western Australia
MONICA TAMARIZ, University of Edinburgh
T MARK ELLISON, University of Western Australia
DALE BARR, University of Glasgow

## 1. INTRODUCTION

Human communication systems, such as language, evolve culturally [Chater et al. 2009; Croft 2000]; their components undergo reproduction and variation. However, a role for selection in cultural evolutionary dynamics is less clear. Often neutral evolution (also known as 'drift') models, are used to explain the evolution of human communication systems [Blythe 2012; Nettle 1999; Reali and Griffiths 2010], and cultural evolution more generally [Bentley et al. 2004; Neiman 1995]. Under this account, cultural change is unbiased: for instance, vocabulary, baby names [Bentley, Hahn and Shennan 2004] and pottery designs [Neiman 1995] have been found to spread through random copying. This is a *neutral* account because all variants encountered are considered equal candidates for copying.

While drift is the null hypothesis for models of cultural evolution [Blythe 2012; Nettle 1999] it does not always adequately explain empirical results [Baxter et al. 2009]. Alternative models include cultural selection, which assumes variant adoption is biased. Theoretical models of human communication argue that during conversation interlocutors are biased to adopt the same labels and other aspects of linguistic representation (including prosody and syntax) [Pickering and Garrod 2004]. This basic alignment mechanism has been extended by computer simulation to account for the emergence of linguistic conventions. When agents are biased to match the linguistic behavior of their interlocutor, a single variant can propagate across an entire population of interacting computer agents [Barr 2004; Steels 2003]. This behavior-matching account operates at the level of the individual. We call it the *Conformity-biased model*. Under a different selection account, called content-biased selection [Boyd and Richerson 1985], functional selection [Nettle 1999] or replicator selection [Baxter, Blythe, Croft and McKane 2009], variant adoption depends upon the intrinsic value of the particular variant (e.g., ease of learning or use). This second alternative account operates at the level of the cultural variant. Following Boyd and Richerson we call it the *Content-biased model*. The present paper tests the drift model and the two biased selection models' ability to explain the spread of communicative signal variants in an experimental micro-society.

### 1.1 Modeled Empirical Data

The present paper models the results of an experimental-semiotic study where human participants communicate a set of fixed concepts by drawing on a shared digital whiteboard [Fay et al. 2010]. In this study participants were not allowed to use conventional language (spoken or written), forcing them to create a new graphical communication system from scratch. Participants were organized into four 8-Person micro-societies, and communicated a list of 16 recurring concepts (e.g., *Art Gallery, Drama, Theater*) to their partner (i.e., all communication took place in pairs). After several games, they switched partners and repeated this process until they had interacted with each of the other members of their group. A representative example of the spread of a cultural variant for the concept *Soap Opera* is given in Figure 1.

What cultural evolutionary dynamics best explain the change in the frequencies of the communication variants illustrated in Figure 1? To answer this question we constructed a probabilistic model that mirrored the structure and pattern of interactions of the experimental micro-societies collected by Fay et al. [2010]. This model has 3 parameters. Memory Size sets the amount of variant history than can affect the an agent's next variant choice: ranging from 2 to 8 in steps of 2. *Conformity-bias* goes from -1 (fully egocentric: preferring self-produced variants over other-produced variants) to 1 (fully allocentric: preferring other-produced variants over self-produced variants) in steps of 0.2. *Content-bias* prefers one variant over all others: ranging from 0 (no preference) to 1 (always adopt biased variant) in steps of 0.1. Mutation probability was fixed at 0.02, consistent with the modelled empirical data. The fit of possible parameter combinations was then assessed against the empirical data. Simulating the behavior of corpus data collected under controlled laboratory conditions minimizes the effect of extraneous variables, and increases our confidence in the explanatory power of the model.

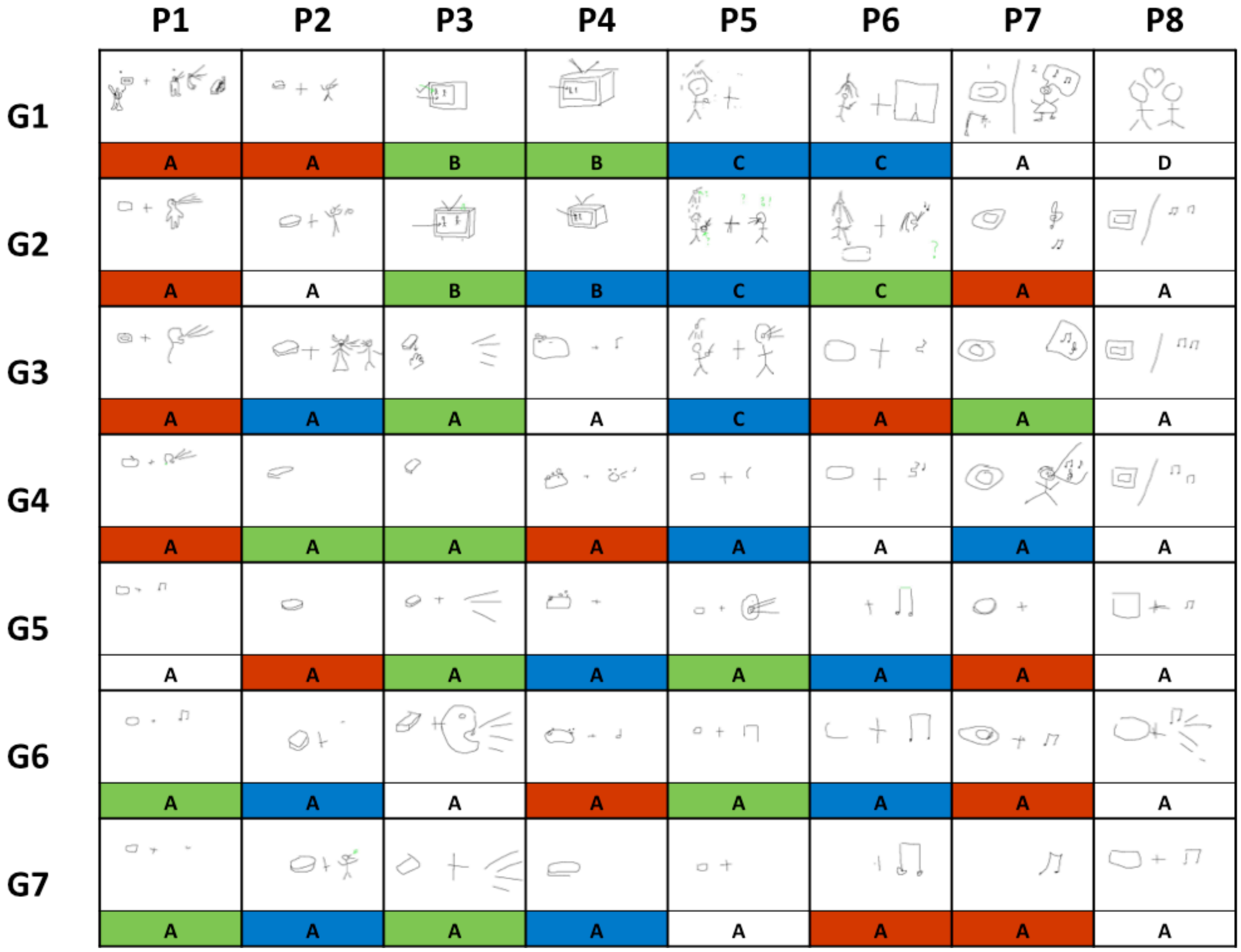

Fig. 1. Cultural evolution of the signs used to represent *Soap Opera* in an 8-Person micro-society [from Fay et al. 2010]. Columns correspond to Participants (P1 to P8) and rows to Generations (G1 to G7). Colors indicate the different participant pairings in a given generation. When participants played with their first partner (Generation G1) they used a variety of different signs: a bar of soap and a musical note (variant A), a television (B), a shower (C) and a love heart (D). In this generation participants tended to adopt the sign their partner produced. As they interacted with the other members of their micro-society, the soap and musical note (variant A) propagates until everyone is using a refined version of this sign (either soap, a musical note or both) by Generation G4. Notice that participants retain their initial variant until they encounter variant A, after which they only use variant A. This suggests a strong Content-bias for variant A. In each micro-society participants communicate 16 concepts, giving 64 distinct data structures like that shown in Figure 1 (a total of 3584 signs).

### 1.2 Model

We constructed a parameterized model of participant variant choice. The model takes as input the history of the representational variants the participant had used or seen a partner use, and returns a distribution over how they might next represent that concept.

Together the model parameters define the 484 points in the parameter space. The likelihood of each of the 64 data-structures was evaluated at each point, and the best parameter setting was retained. The strength of evidence for a bias in particular data structures was evaluated using a best-account Bayes' Factor: the maximum likelihood of any model with the bias divided by the maximum likelihood of any model without the bias. This approach is formally equivalent to Kass and Raftery [1995]'s use of Bayes' factor, although the thresholds for different strengths of support differ slightly. Whereas they count *strong support* from a Bayes Factor of 20, our threshold for *significant evidence* (in keeping with the standard p<.05 significance criterion) is 19.

### 1.3 Results and Discussion

Lower Memory Size (2 or 4) was associated with better model fit. Contrary to a conformity bias, an egocentric bias (-1.0 to -0.5), where agents tend to reuse variants they have used previously, was associated with better model fit. Most data structures are best accounted for with some Content-bias (95% of data structures). Whereas 28% of the data-structures are consistent with a drift account, 72% are consistent with a biased account (Conformity and Content; see Fig. 2).

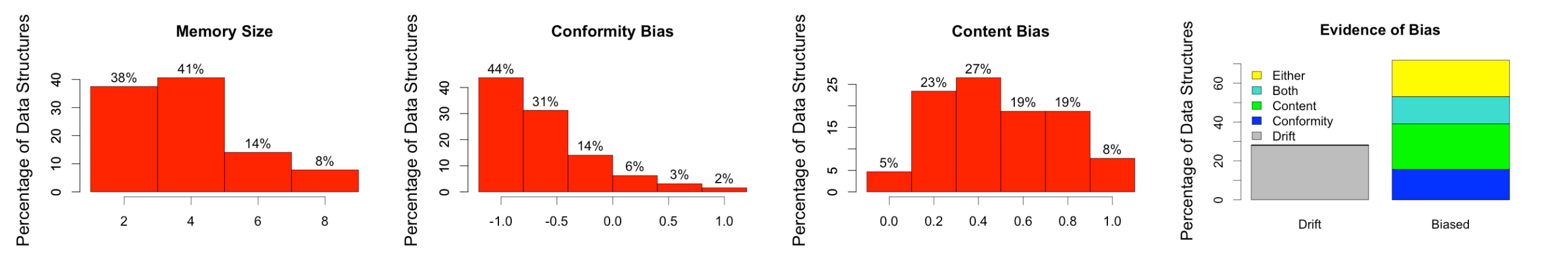

| (a) | (b) | (c) | (d) |
|---|---|---|---|

Fig. 2. Histograms showing the frequency of different Memory Sizes (a), Conformity-bias (b), and Content-bias (c) in the best-fit models for the 64 data-structures. Panel (d) shows the number of data-structures that are explained by the biases. These break down into four types: 23% show significant evidence for Content-bias but not Conformity bias (green), 16% show conformity bias without Content-bias (blue), 14% have significant evidence for both (turquoise), and the remaining 19% show significant evidence for the presence of some bias, but there is insufficient evidence to pinpoint which biases are active (yellow).

The present study extends neutral models to show that selection plays an important role in the cultural evolution of human communication systems. Whereas the median Bayes' Factor for Conformity-bias alone and Content-bias alone is below the significant evidence criterion of 19 (6.03 and 14.11 respectively), together they returned a median value of 71.52. This indicates a critical interplay between the biases: people tend to reuse variants they have used in the past unless the newly encountered variant is superior, in which case it is adopted (because the content bias typically overwhelms the egocentric bias). Consistent with our findings in support of content bias, empirical studies have shown, using the same corpus from Fay et al. [2010], that the signs selected by the laboratory micro-societies confer distinct learning and usability benefits [Fay and Ellison 2013; Fay et al. 2008]. The present empirically-grounded model challenges simluations in which population-level sign propogation relies on alignment and reinforcement learning [Barr 2004; Steels 2003]. Accepting that content bias is a driver behind the spread of communication variants supports the view that human communication systems, such as language, are functionally adaptive complex systems [Beckner et al. 2009].